\documentclass{fairmeta}
\usepackage{amsmath}
\usepackage{enumerate} 
\usepackage{bm}
\usepackage{algorithm}
\usepackage{floatflt}
\usepackage{algpseudocode}
\usepackage{amsfonts}
\usepackage{amsthm}
\usepackage{newtxtt}
\usepackage{algorithm}
\usepackage{colortbl} 
\usepackage{cleveref}
\usepackage{diagbox} 
\usepackage[utf8]{inputenc}
\usepackage{textgreek}
\usepackage{colortbl}
\usepackage{nicematrix}
\usepackage{makecell}
\usepackage{float}
\usepackage{arydshln}
\usepackage[frozencache,cachedir=.]{minted}
\usepackage{caption}
\usepackage{subcaption}
\usepackage{tcolorbox}
\usepackage{amssymb}
\usepackage{xspace}
\usepackage{wrapfig}
\usepackage{adjustbox}
\usepackage{tabularx}
\usepackage{booktabs}
\usepackage{mathtools}
\usepackage{wrapfig}
\usepackage{amssymb}
\usepackage{graphicx}

\usepackage{silence}
\makeatletter
\patchcmd{\wrong@fontshape}{\@gobbletwo}{}{}{}
\makeatother
\definecolor{upColor}{RGB}{17,138,21}
\definecolor{downColor}{RGB}{174,36,67}

\newtheorem{theorem}{Theorem}[]

\newtheorem{remark1}[theorem]{Remark}

\title{Enhancing Neural Video Compression of Static Scenes with Positive-Incentive Noise}
\author[]{Cheng Yuan}
\author[]{Zhenyu Jia}
\author[]{Jiawei Shao}
\author[]{Xuelong Li}
\affiliation[]{Institute of Artificial Intelligence (TeleAI), China Telecom}
\metadata[Correspondence to]{Xuelong Li (\email{xuelong\_li@ieee.org})}

\begin{document}

\abstract{
Static scene videos, such as surveillance feeds and videotelephony streams, constitute a dominant share of storage consumption and network traffic. However, both traditional standardized codecs and neural video compression (NVC) methods struggle to encode these videos efficiently due to inadequate usage of temporal redundancy and severe distribution gaps between training and test data, respectively. While recent generative compression methods improve perceptual quality, they introduce hallucinated details that are unacceptable in authenticity-critical applications. To overcome these limitations, we propose a positive-incentive camera (PIC) framework for static scene videos, where short-term temporal changes are reinterpreted as positive-incentive noise to facilitate NVC model finetuning. By disentangling transient variations from the persistent background, structured prior information is internalized in the compression model. During inference, the invariant component requires minimal signaling, thus reducing data transmission while maintaining pixel-level fidelity. Experiment results show that PIC achieves visually lossless reconstruction for static scenes at an extremely low compression rate of 0.009\%, while the DCVC-FM baseline requires 20.5\% higher Bjøntegaard delta (BD) rate. Our method provides an effective solution to trade computation for bandwidth, enabling robust video transmission under adverse network conditions and economic long-term retention of surveillance footage.
}

\maketitle

\section{Introduction}

With the expanding adoption of visual sensing and real-time communication, static scene videos, exemplified by surveillance feeds and videotelephony streams, constitute a dominant portion of storage consumption and network traffic.
These videos typically contain limited and localized temporal variations with nearly stationary backgrounds, yet their prolonged duration results in immense data volumes, posing significant challenges to existing compression techniques.
Traditional standardized video codecs, such as H.264/AVC \citep{H264}, H.265/HEVC \citep{H265}, and H.266/VVC \citep{H266}, exhibit insufficient coding efficiency for static scene videos, as their hand-crafted motion compensation and transform coding pipelines fail to fully exploit the temporal redundancy and structural regularity inherent in these videos.

To address this limitation, neural video compression (NVC) has achieved superior rate-distortion performance by leveraging end-to-end optimization and learned entropy models \citep{ssf2020, liu2024bidirectional, DCVC-FM, DCVC-RT}.
However, their efficiency degrades drastically when deployed on static scene videos, primarily due to the severe distribution gap between the dynamic-rich training data and the static-dominant testing videos.
This discrepancy leads to suboptimal motion modeling and inefficient bit allocation.
Inspired by recent advances in video generation models, generative methods are employed to synthesize high-frequency details, thus eliminating compression artifacts and enhancing perceptual quality \citep{GVC_teleai_2507, GVC_teleai_2512, GLC-video}.
Unfortunately, these approaches often introduce artificial textures and hallucinated details that are intolerable in critical scenarios such as surveillance and video conferencing, where pixel-level fidelity and content authenticity are mandatory requirements.

Beyond video compression, \textbf{positive-incentive noise} has demonstrated remarkable effectiveness across multiple machine learning tasks \citep{PiNoise_li2022, PiNoise_zhang2025}.
By deliberately injecting structured perturbations during representation learning, positive-incentive noise encourages models to learn more discriminative features, thus improving robustness and generalization capability.
The efficacy of this paradigm has been validated in diverse domains, including but not limited to visual-language alignment \citep{PiNoise_VL_zhang2025} 
, graph learning \citep{PiNoise_graph_zhang2025, PiNoise_graph_contrastive_zhang2025}, unsupervised clustering \citep{PiNoise_cluster_chen2024}, and class-incremental learning \citep{PiNoise_continual_learning}.

Motivated by these successes, we propose a \textbf{positive-incentive camera} (PIC) framework for static scene videos, where short-term temporal changes, such as object movements or light flickerings, are treated as positive-incentive noise when training an NVC model.
These dynamic elements incentivize the model to disentangle transient variations from persistent background representations, from which the structured prior information is extracted.
With the assistance of positive-incentive noise, the prior knowledge can be progressively learned and implicitly stored in the compression model through on-device finetuning.
During inference, the learned background prior requires minimal signaling, and the dynamic variations account for the primary components in the encoded and transmitted symbols.
Our framework leverages positive-incentive noise as an instructive training mechanism instead of a generative solution, thus enhancing rate-distortion (RD) performance without compromising reconstruction authenticity.

The proposed PIC framework aligns with the rapid development of artificial intelligence (AI) services deployed in communication and computation integrated networks, exemplified by the three-tier architecture described in the \textbf{AI Flow} framework \citep{aiflow, aiflow_perspectives}.
Contrary to the mainstream cloud-centric computing scheme, these systems coordinate device, edge, and cloud resources to support latency-sensitive applications even in the most challenging circumstances, as explored by task-oriented communication for computer vision \citep{shao2021learning, shao2024task, TOFC} and the law of information capacity in large language models \citep{yuan2026informationcapacityevaluatingefficiency}.
The soaring capabilities of edge hardware have opened up the possibility of \textbf{trading computation for bandwidth}, where neural networks are employed on user devices to extract compact representations and minimize data transmission \citep{fan2025computationbandwidthmemorytradeoffsunifiedparadigm}.
This trade-off exhibits particular significance when channel conditions fluctuate unpredictably or when spectrum restrictions become stringent due to high concurrency.
Our method extends this idea to the scenario of static scene video compression and provides a robust solution to deliver smooth and high-definition video streams in adverse network environments.
Apart from real-time communication, PIC also substantially reduces the storage costs of surveillance footage, rendering data retention over an extended period more economically viable.

\section{Results}
\subsection{Implementation Details}
We have recorded a total of over 132 hours of static scene surveillance videos in three different scenarios with a resolution of $2560 \times 1440$ and a frame rate of 25.
The proposed PIC framework is employed to finetune the deep contextual video coding with feature modulation (DCVC-FM) model \citep{DCVC-FM} on a 60-hour subset of this dataset.
During training, a 32-frame clip is randomly selected from each video sample with a resolution of $1792 \times 1024$, and the initial learning rate is set to $10^{-6}$.
The distortion loss is measured by mean squared error (MSE) in the YUV color space, with the Y channel given a weighting coefficient six times larger than that of other channels.
The Lagrange multiplier that controls the RD trade-off ranges from $2 \times 10^{-5}$ to 0.0018 as the quality parameter changes from 0 to 63.
The remaining hyperparameters and implementations are identical to those of the CompressAI library \citep{compressai} and the official DCVC-FM codebase \citep{DCVC-FM}, including the Adam optimizer, the reduce-on-plateau learning rate scheduling, using the quality parameter offsets $[0, 1, 0, 2, 0, 2, 0, 2]$ for each 8-frame group, and resetting internal reference features every 32 frames to ensure long-term stability.

\subsection{RD Performance}
We evaluate the RD performance on randomly selected 600-frame clips outside the subset used for finetuning.
These clips are divided into two groups, namely static and dynamic, based on the intensity of inter-frame changes.
The data size is measured by the average bits per pixel (BPP) of all frames, and the reconstruction quality is assessed by the average peak signal-to-noise ratio (PSNR) in the YUV color space.
The PSNR value of the Y channel is given a weighting coefficient six times larger than that of other channels, following the official implementation of DCVC-FM evaluations \citep{DCVC-FM}.
We use a widely accepted metric, Bjøntegaard delta (BD) rate, to facilitate holistic evaluation, which calculates the ratio of data sizes required by two compression methods to achieve a specific reconstruction quality, averaged across the PSNR range that is achievable for both methods.

\begin{figure}[htb]
    \centering
    \includegraphics[width=.8\linewidth]{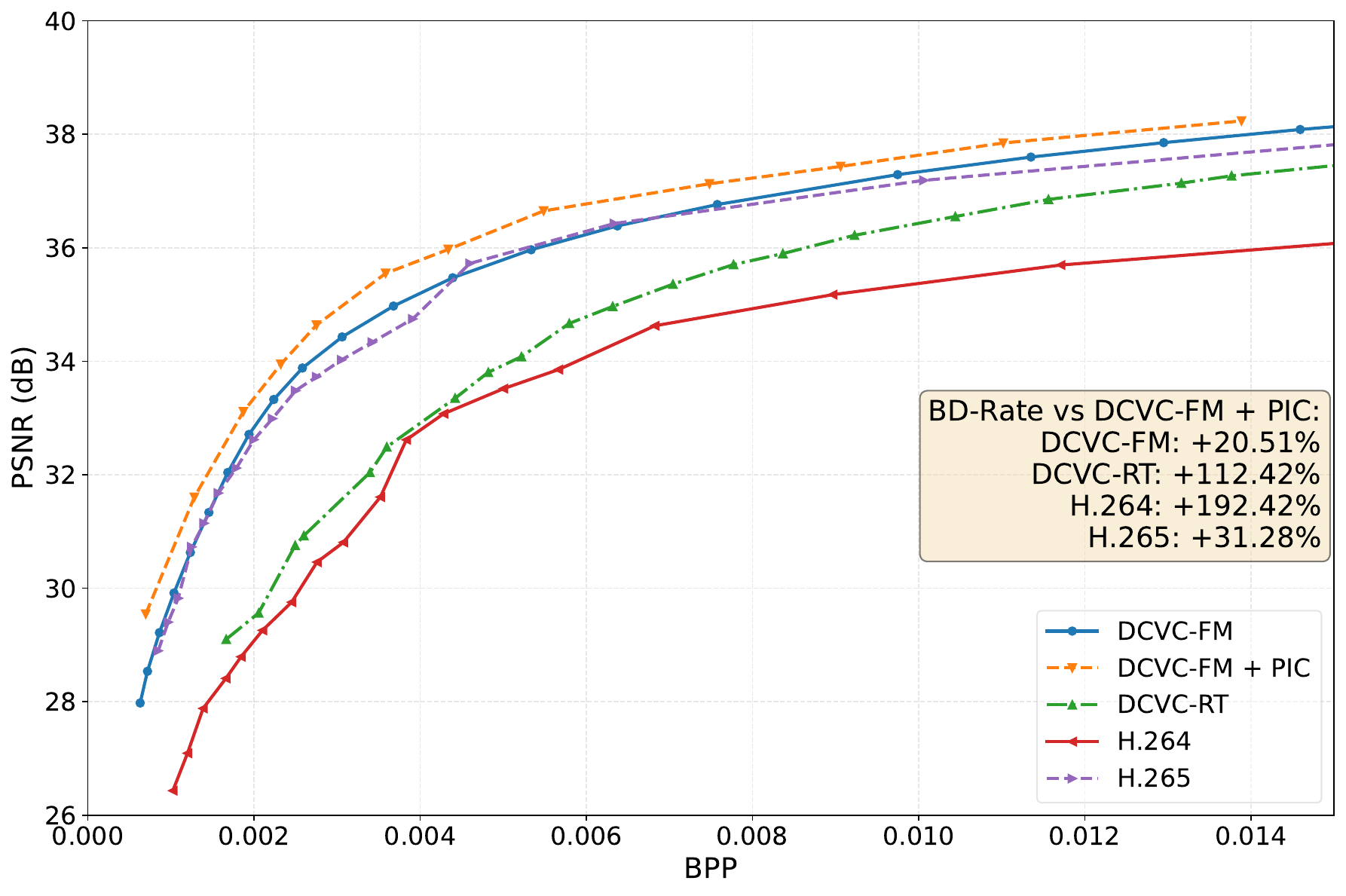}
    \caption{\textbf{RD performance comparison of different compression methods on videos in the static group.} The DCVC-FM requires 20.5\% more BD-rate compared to the PIC finetuned variant, demonstrating the efficacy of positive-incentive noise in optimizing NVC models. Traditional H.265 and H.264 codecs require 31.3\% and 192\% higher BD-rate, respectively.}
    \label{fig:dcvc_RD}
\end{figure}

As illustrated in Figure~\ref{fig:dcvc_RD}, our PIC approach delivers superior RD performance on videos in the static group, compared to other neural codecs (DCVC-FM \citep{DCVC-FM} and DCVC-RT \citep{DCVC-RT}) and traditional standards (H.264 and H.265).
The proposed PIC finetuning brings consistent performance improvements to the DCVC-FM model across a wide PSNR range from 29 to 38 dB.
In the extremely low BPP range of less than 0.002, DCVC-FM fails to maintain the performance advantage over H.265.
Nevertheless, PIC still provides significant data size savings beyond this remarkable compression rate of 0.008\%.
On average, DCVC-FM requires 20.5\% more data to achieve identical PSNR qualities compared to the PIC finetuned variant, demonstrating the efficacy of positive-incentive noise in optimizing NVC models.
DCVC-RT delivers a 112\% higher BD-rate, as the simplified network structure without explicit motion estimation degrades the model's efficiency to encode static background information.
Traditional H.265 and H.264 codecs require 31.3\% and 192\% higher BD-rate, respectively, inferior to DCVC-FM counterparts.

\begin{figure}[htb]
    \centering
    \includegraphics[width=.8\linewidth]{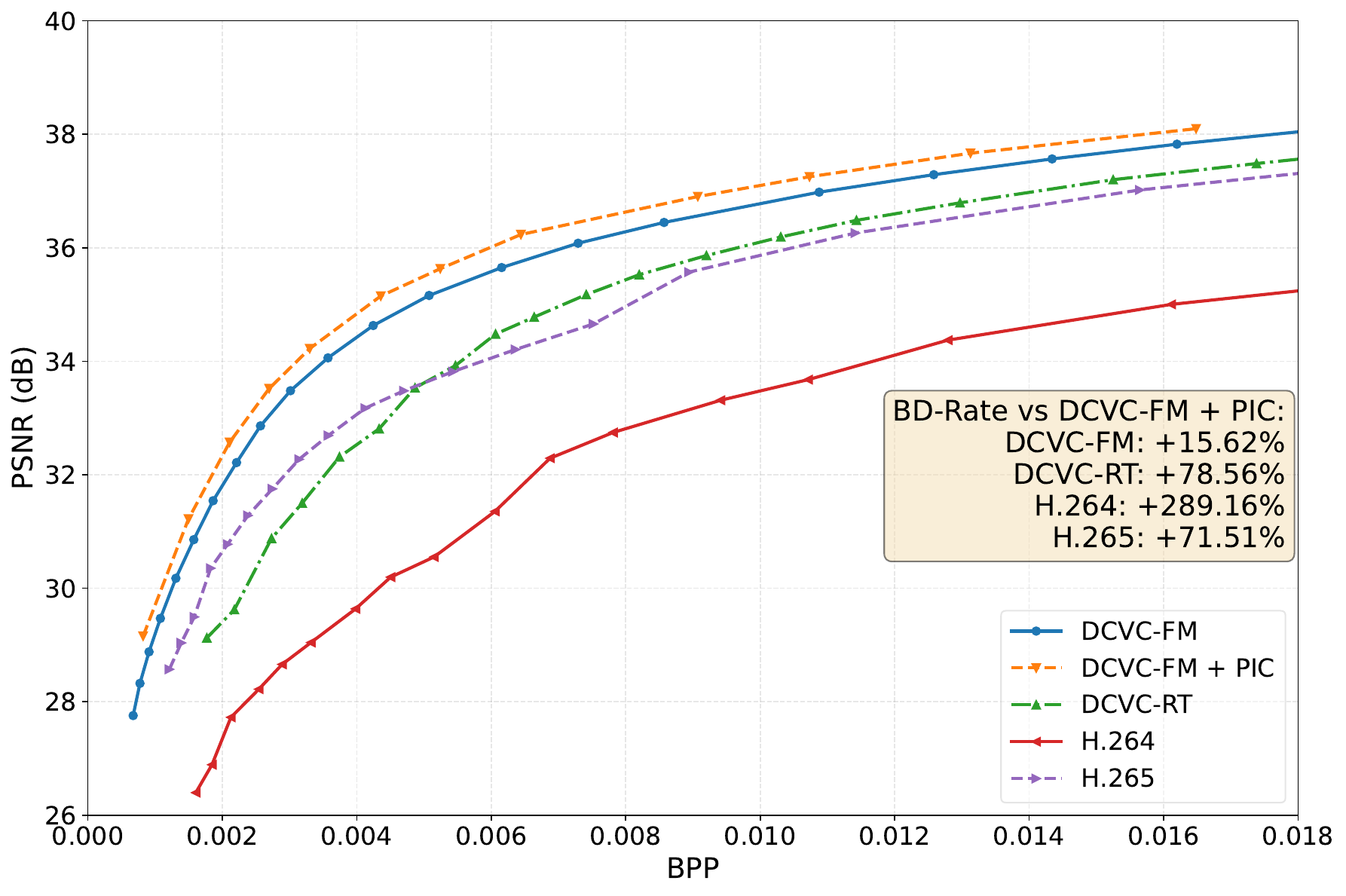}
    \caption{\textbf{RD performance comparison of different compression methods on videos in the dynamic group.} DCVC-FM requires 15.6\% more BD-rate compared to the PIC finetuned variant, validating the benefits of positive-incentive noise even when the distribution gap is narrower. Standard H.265 and H.264 codecs incur 71.5\% and 289\% BD-rate penalties, respectively.}
    \label{fig:dcvc_RD_dynamic}
\end{figure}

As for videos with more temporal variations, Figure~\ref{fig:dcvc_RD_dynamic} compares the RD performance of different compression methods.
DCVC-FM requires 15.6\% more BD-rate compared to the PIC finetuned variant, validating the benefits of positive-incentive noise even when the distribution gap in video sources is narrower.
Standard H.265 and H.264 codecs incur 71.5\% and 289\% BD-rate penalties, respectively, significantly higher compared to the previous case with more static videos.
To sum up, the proposed PIC allows NVC models to achieve superior efficiency over traditional codecs for typical static scene videos with minor inter-frame changes, while increasing the performance advantage of NVC models for the less frequent case with pronounced temporal variations.

\subsection{Visual Comparison}
\begin{figure*}[htb]
    \centering
    \includegraphics[width=\linewidth]{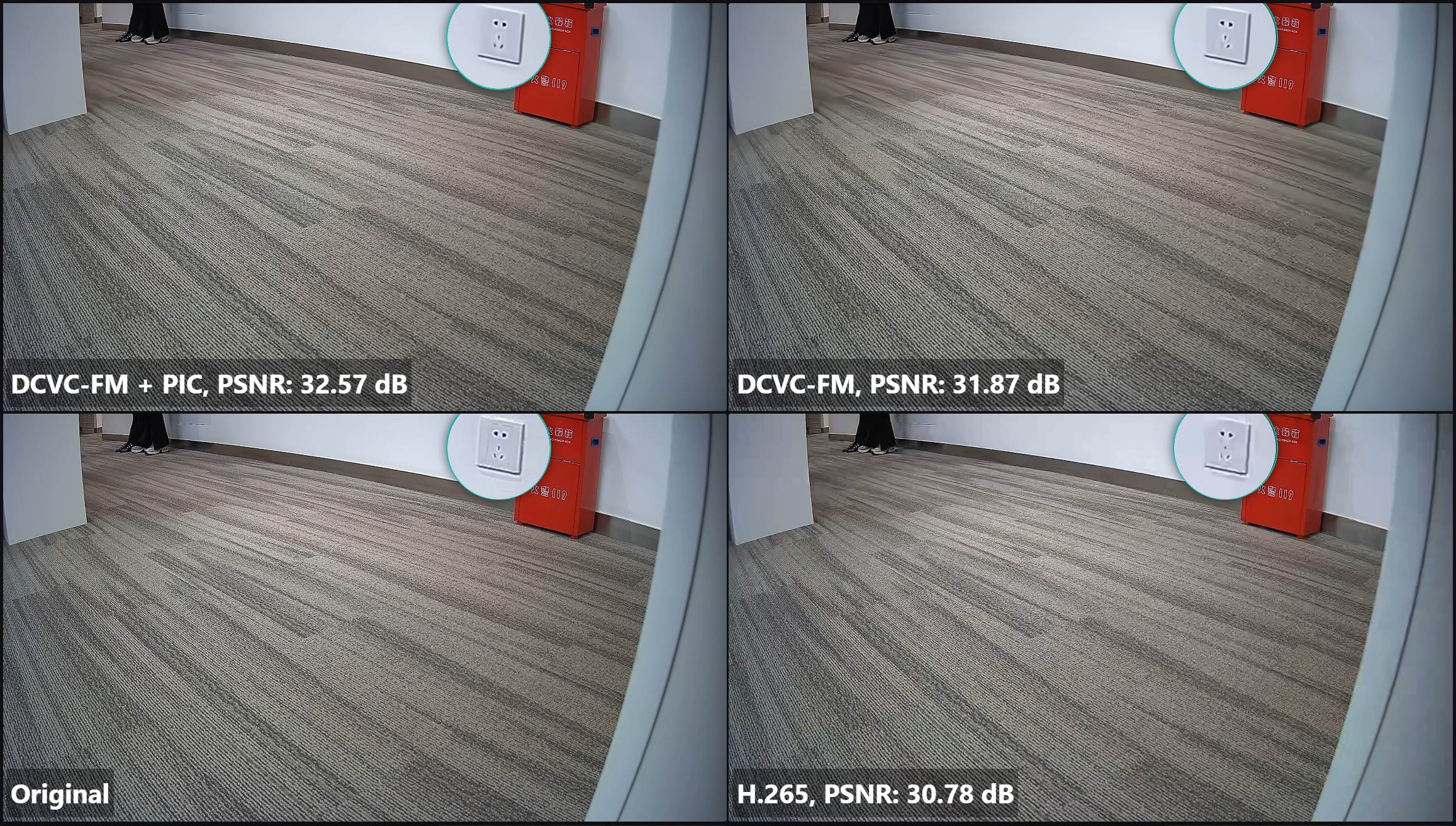}
    \caption{\textbf{Visual comparison of video frames reconstructed by different compression methods at a BPP of 0.002.} The proposed PIC approach maintains the details of the wall socket that the DCVC-FM baseline fails to reconstruct, and avoids the blurring and color drifting impairments caused by H.265 compression. PIC achieves visually lossless reconstruction for static scenes at an extremely low compression rate of 0.009\%.}
    \label{fig:dcvc_visual_compare}
\end{figure*}

Figure~\ref{fig:dcvc_visual_compare} provides a visual comparison of video frames reconstructed by the traditional codec H.265 and DCVC-FM models before and after PIC finetuning.
To ensure a fair comparison, we carefully select the hyperparameters of H.265 and the quality parameters of DCVC-FM models to maintain nearly identical data rates of approximately 0.002 BPP.
The proposed approach based on positive-incentive noise maintains the details of the wall socket that the DCVC-FM baseline fails to reconstruct, and avoids the blurring and color drift artifacts caused by H.265 compression.
Quantitative evaluation shows a PSNR improvement from 31.87 dB to 32.57 dB under nearly identical data rates, far exceeding the traditional H.265 codec's 30.78 dB.
PIC achieves visually lossless reconstruction for static scenes at an extremely low compression rate of 0.009\%.
For videos with more inter-frame changes, the required bitrate is higher to eliminate perceptible blurring around moving objects, which is not the dominant case of static scene videos.

\subsection{Extension to Other NVC Models}
We extend the PIC framework to the space-scale flow (SSF) video compression model \citep{ssf2020} to demonstrate its generalization capability.
Five checkpoints released by the CompressAI package \citep{compressai} are utilized as initial weights, which correspond to quality levels of 1, 4, 6, 7, 8, constituting the upper envelope of the RD performance curve before finetuning.
During training, a 20-frame clip is randomly selected from each video sample with a resolution of $2560 \times 1408$ to accommodate the network requirements, and the initial learning rate is set to $2 \times 10^{-5}$, with an individual optimizer with a constant learning rate of $10^{-3}$ for the quantiles in the learned entropy models.
The values of the Lagrange multiplier are identical to the baseline models, given by 0.0018, 0.013, 0.0483, 0.0932, and 0.18, respectively.

\begin{figure}[!htb]
    \centering
    \includegraphics[width=.8\linewidth]{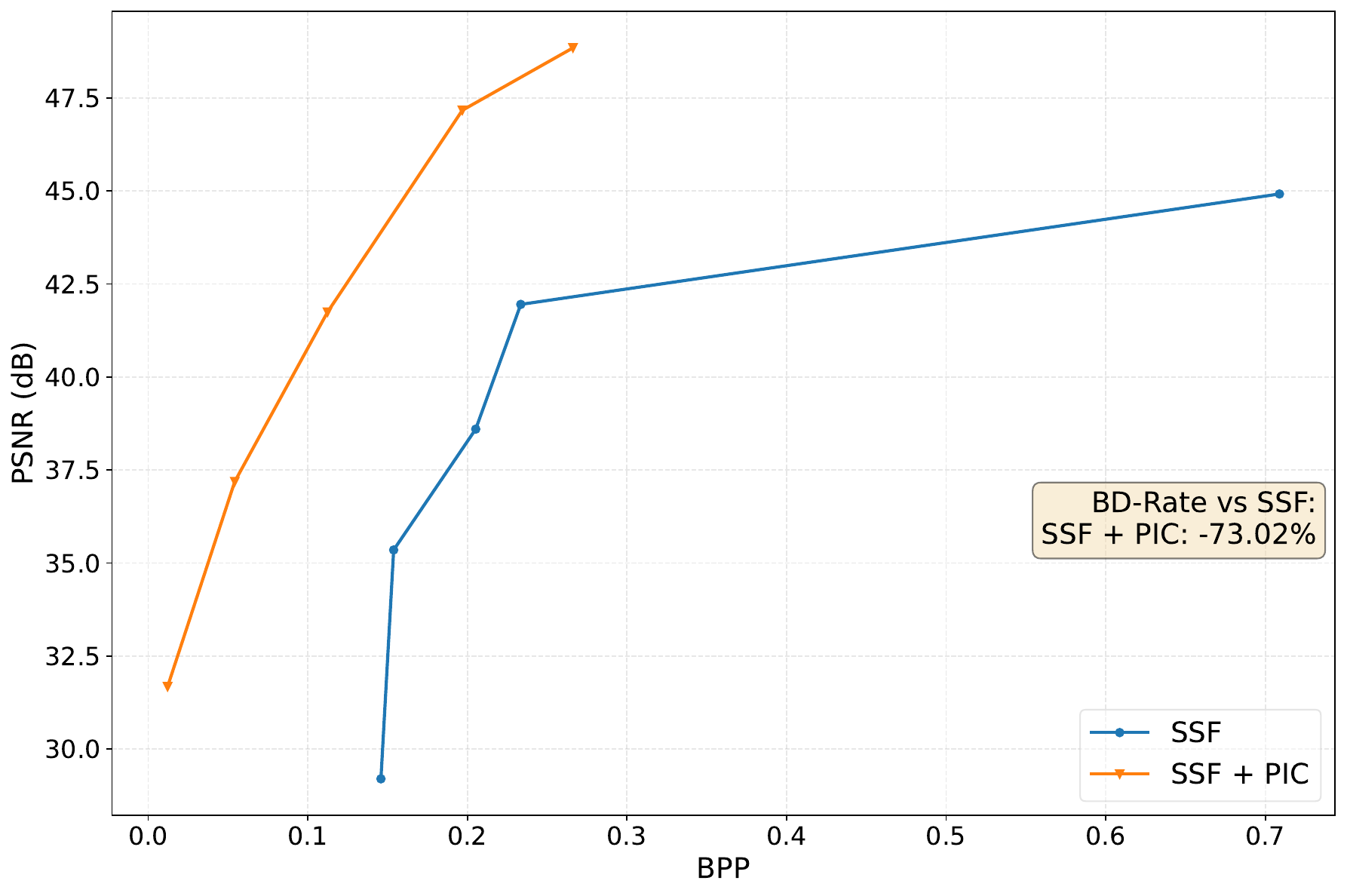}
    \caption{\textbf{RD performance comparison of SSF models before and after PIC finetuning.} The proposed PIC approach achieves a significant BD-rate reduction of 73\%, demonstrating the efficacy of positive-incentive noise in optimizing NVC models.}
    \label{fig:ssf_RD}
\end{figure}

As shown in Figure~\ref{fig:ssf_RD}, PIC simultaneously increases the PSNR quality and reduces the data size for all five checkpoints, demonstrating the efficacy of positive-incentive noise in training video compression models.
To achieve a PSNR quality lower than 42.5 dB, the proposed method can decrease the required BPP by at least 0.1, more than half of the original data volume.
In a higher PSNR range, the rate savings are even more prominent, and the highest achievable PSNR is further increased from 44.92 dB to 48.84 dB while consuming only 37.6\% of the original storage space.
Overall, a 73.0\% Bjøntegaard delta (BD) rate saving is achieved in comparison to original checkpoints, measured with cubic spline interpolation across a wide PSNR range from 32 dB to 45 dB.

\begin{figure*}[!htb]
    \centering
    \includegraphics[width=\linewidth]{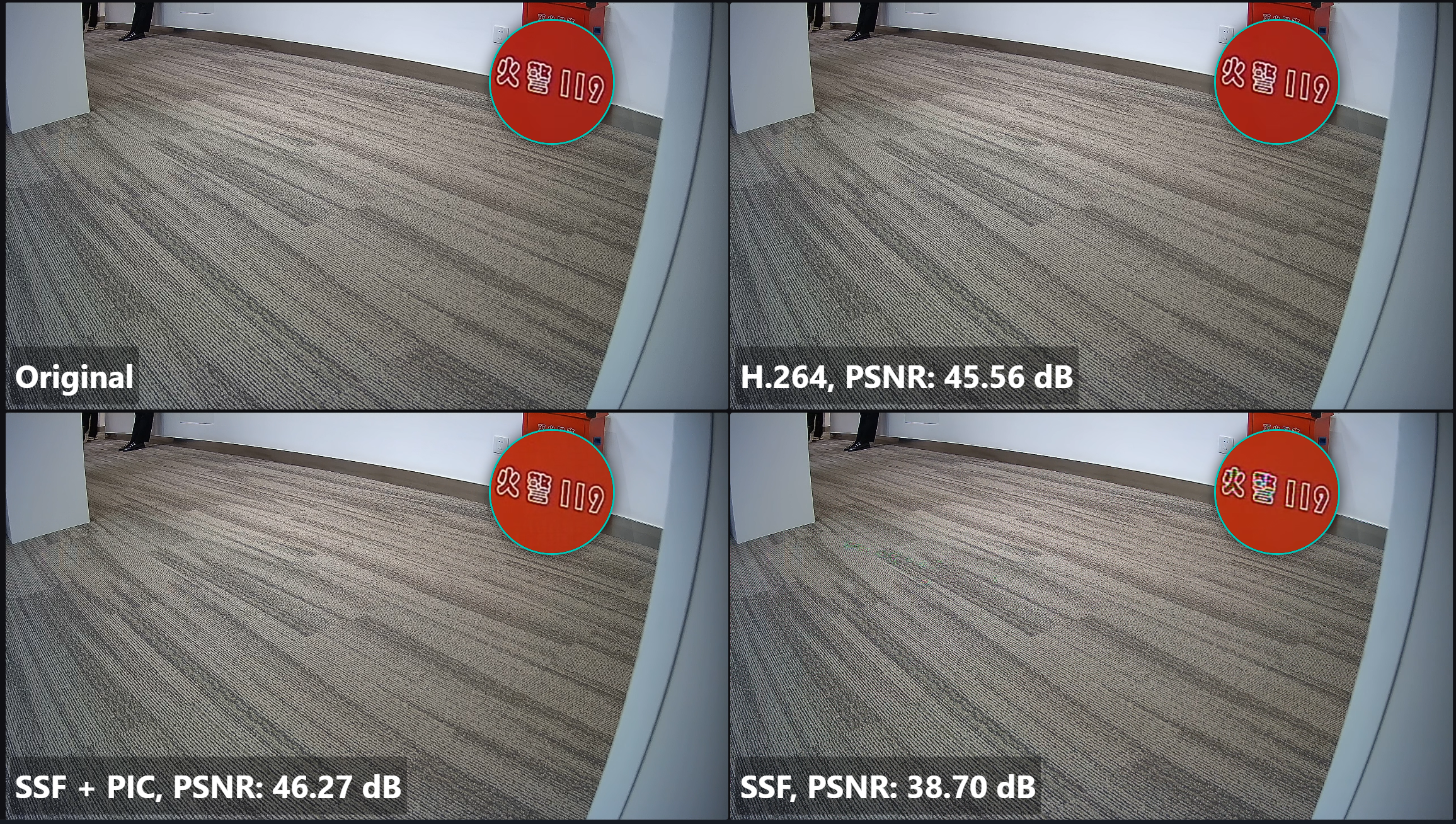}
    \caption{\textbf{Visual comparison of video frames reconstructed by different compression methods at a BPP of 0.2.} The proposed PIC approach enhances the PSNR quality from 38.70 dB to 46.27 dB under nearly identical data rates.}
    \label{fig:ssf_visual_compare}
\end{figure*}

Figure~\ref{fig:ssf_visual_compare} provides a visual comparison of video frames reconstructed by the traditional codec H.264 and SSF models before and after PIC finetuning.
To ensure a fair comparison, we carefully select the hyperparameters of H.264 and the checkpoints of SSF models to maintain nearly identical data rates of approximately 0.2 BPP.
The proposed approach based on positive-incentive noise enhances the PSNR quality from 38.70 dB to 46.27 dB with a slight reduction in BPP from 0.205 to 0.197.
Direct visual comparison verifies the improved reconstruction quality, where the baseline model produces noticeable colored fringes on the text of the fire extinguisher container at this bitrate, while our finetuned model successfully preserves the original clarity.
The actual reconstruction quality of the SSF model evaluated on our static scene videos is substantially lower than that of the traditional H.264 codec.
This result is contrary to the original paper \citep{ssf2020}, which shows consistent performance gains over H.264 when BPP exceeds 0.1, mainly due to the distribution gap between the training dataset and static scene videos used in our evaluation.
Guided by positive-incentive noise, our online finetuning method successfully bridges this gap and achieves superior performance compared to H.264.

\section{Conclusion}
This report proposes a PIC framework to enhance the efficiency of NVC for static scene videos, where transient temporal changes serve as positive-incentive noise that benefits model finetuning.
These variations stimulate the model to learn robust prior information from the persistent background, which is internalized in model parameters and requires minimal data transmission during inference.
This method represents a new way of trading computation for bandwidth in communication and computation-integrated networks, filling the vacancy in authenticity-critical scenarios left by previous generative video compression methods.
Experiment results show that PIC achieves visually lossless reconstruction for static scenes at an extremely low compression rate of 0.009\%, while the DCVC-FM baseline requires 20.5\% higher Bjøntegaard delta (BD) rate.
The proposed approach enables economic long-term retention of surveillance footage and robust video streaming against network fluctuations.
In the near future, we will further enrich the experiments and provide concrete theoretical foundations on the role of positive-incentive noise.

\bibliographystyle{plainnat}
\bibliography{paper}

\end{document}